\documentclass[12pt,preprint]{aastex}
\begin{document}

\title{Binary Central Stars of Planetary Nebulae Discovered Through Photometric Variability III: The Central Star of Abell~65}
\shorttitle{The Central Star of Abell 65}
\author{Todd C. Hillwig\altaffilmark{1,11}}
\altaffiltext{1}{Department of Physics and Astronomy, Valparaiso University, Valparaiso, IN 46383 USA and SARA}
\email{Todd.Hillwig@valpo.edu}
\author{David J. Frew\altaffilmark{2,3}}
\altaffiltext{2}{Department of Physics, The University of Hong Kong, Pokfulam Road, Hong Kong}
\altaffiltext{3}{Department of Physics and Astronomy, Macquarie University, Sydney NSW 2109, Australia}
\author{Melissa Louie\altaffilmark{4,9}}
\altaffiltext{4}{Department of Physics \& Astronomy, Stony Brook University, Stony Brook, NY 11794 USA}
\author{Orsola De Marco\altaffilmark{3}}
\author{Howard E. Bond\altaffilmark{5,6,11}}
\altaffiltext{5}{Department of Astronomy \& Astrophysics, Pennsylvania State University, University Park, PA 16802 USA }
\altaffiltext{6}{Space Telescope Science Institute, 3700 San Martin Dr., Baltimore, MD 21218}
\author{David Jones\altaffilmark{7,8}}
\altaffiltext{7}{Instituto de Astrof\'isica de Canarias, E-38205 La Laguna, Tenerife, Spain}
\altaffiltext{8}{Departamento de Astrof\'isica, Universidad de La Laguna, E-38206 La Laguna, Tenerife, Spain}
\author{S. C. Schaub\altaffilmark{1,9,10}}
\altaffiltext{9}{Southeastern Association for Research in Astronomy (SARA) NSF-REU Summer Intern}
\altaffiltext{10}{Current address: Plasma Science \& Fusion Center, Massachusetts Institute of Technology, Cambridge, MA 02139, USA}
\altaffiltext{11}{Visiting astronomer, Cerro Tololo Inter-American Observatory, National Optical Astronomy Observatory, which are operated by the Association of Universities for Research in Astronomy, under contract with the National Science Foundation.}

\begin{abstract}

A growing number of close binary stars are being discovered among central stars of planetary nebulae.  Recent
and ongoing surveys are finding new systems and contributing to our knowledge of the evolution of close binary
systems.  The push to find more systems was largely based on early discoveries which suggested
that 10 to 15\% of all central stars are close binaries.  One goal of this series of papers is confirmation and classification of these systems as close binaries and determination of binary system parameters.  Here we provide time-resolved multi-wavelength photometry of the central star of Abell 65 as well as further analysis of the nebula and discussion of possible binary--nebula connections.  Our results for Abell 65 confirm
recent work showing that it has a close, cool binary companion, though several of our model parameters disagree with the recently published values.  With our longer time baseline of photometric observations from 1989--2009 we also provide a more
precise orbital period of 1.0037577 days. 

\end{abstract}
\keywords{binaries: close --- planetary nebulae: individual (Abell~65)}

\section{INTRODUCTION}

The complex shapes of planetary nebulae (PNe) provide a valuable and narrow evolutionary window through which we can
study the life cycles of intermediate mass stars (typically 1--8 M$_\odot$).  Binary interactions are known to play an important
role but the extent to which such interactions are needed to make a non-spherical PN is still unclear \citep[for a review see][]
{dem09}.  Studies have shown that
approximately ten to twenty percent of all PNe harbor {\it close} binary stars with orbital periods typically less than 3 days
\citep{bon00,mis09a}.  Such close binaries are the result of a common envelope (CE) interaction \citep{pac76}.
The sample of known close binary central stars of planetary nebulae (CSPNe)
more than doubled with the work of \citet{mis09a}.  However, stellar and system parameters for these objects are very
poorly known \citep*[][Papers I \& II]{dem08,hil10}.  Until this situation is alleviated we cannot study this
phenomenon statistically.  \citet{mis09b} use this close binary sample to conclude that the nebular
morphologies of PNe with close binary central stars tend to be bipolar \citep[see also][]{bon90}.  Also, \citet{hil10b}
provides a first look at the binary parameters of the known systems and the relationship to similar systems
which have no observed PN (either because they are more evolved, or because they did not form a visible PN at all).
Additionally, there is increasing observational evidence for a class of intermediate-period CSPN binaries (with periods of months to years).  These binaries are discovered directly from radial velocity studies \citep[e.g.][]{van14}, or otherwise inferred via evidence of mass transfer from the PN's progenitor star onto a companion \citep[e.g.][]{bon03, mis12, mis13b, tyn13}.  For a review, see \citet{van14}.

Planetary nebulae with post CE binaries are also extremely useful as probes of the CE interaction. Not only do they provide us with recent post CE binaries, where the period has not had time to evolve further, but they also show us the ejected envelope itself. As such, they are extremely useful in understanding the CE efficiency \citep{dem11}, but also in the timing of the various CE phases and the energetics of the ejecta \citep{toc14}. For the latter purpose we need to know not only the details of the central binary, but also of the surrounding nebula \citep[see for instance,][]{jon14}

One of the goals of this series of papers is to further our understanding of the binary system parameters for known binary
CSPNe.  Connections between the binary parameters and the surrounding PN can then be explored, whether on a
system-by-system basis, or through a statistical analysis of the larger sample.  The former can be done as each model is
produced.  The latter depends on relatively few objects which currently have parameters derived from system modeling.
We present here a study of the binary central star of Abell 65, with a discussion of the surrounding PN and links
between the two.

\section {BACKGROUND}

Abell 65 (PN G017.3-21.9) was first discovered by \citet{sha59} and independently catalogued as a PN by \citet{abe66}. 
The 15th-mag central star was identified by Abell, who presented $UBV$ photometry. Curiously, Abell did not note variability of the star, based on 4 separate observations, even though he did report variability for several other nuclei which was confirmed later. The variations of the Abell~65 nucleus were first discovered by H.E.B. in 1989. It soon became clear that the star was a close binary with an orbital period very close to 1~day, as was reported by \citet{bon90}. The $\sim$1 day period probably explains why the large-amplitude variability was not detected by Abell: the orbital phase will be nearly the same at the same sidereal time every night during a short observing run! 
A later paper by \citet{wal96} erroneously classified the CS as an eclipsing binary.
\citet{wal96} also suggested that the CS was a cataclysmic variable (CV) star based on strong Balmer emission lines observed
in the spectrum.  Improved light curves \citep{shi09,lou10} along with phase-resolved stellar spectra \citep{shi09}
have since confirmed that the photometric variability is due to a strong irradiation effect, rather than eclipses, and that the
irradiation effect is also the source of the strong emission lines observed in the optical spectrum.  Those papers also
confirmed a period very close to one day.

Physical values for the nebula such as distance, size, and age are given by a number of authors \citep[e.g.][]{wal96,phi05,fre08,fre14}.
\citet{huc13} also provide a three-dimensional model of the nebula.  They show a clearly bipolar nebula with both inner
and outer structures \citep[the outer shell was first identified by][]{lon77}.  As noted by \citet[][based on an image from \citet{hua98}]{mis09b}, \citet{huc13} confirm that while the bright nebular structure may resemble a toroidal
waist, it is actually a set of bipolar lobes with fainter outer lobes (see their Figure 5).
The inclination of the symmetry axis of these structures in their model are $68\pm10\degr$ and $55\pm10\degr$ for the
inner and outer lobes, respectively.  They also calculate kinematic ages for the inner and outer shells respectively, dependent on the distance to the nebula, of $(8000\pm3000) (d/{1\,\rm kpc})$ yr and $15,000\pm5000 (d/{1\,\rm kpc})$ yr, where $d$ is the distance of the PN.  Distances to the nebula range from $d=1.17$ kpc from \citet{fre08} to $d = 1.67$ kpc from \citep*{sta08}.  Hereafter we adopt the updated distance of \citet{fre14}, $d = 1.4$ kpc with a reported uncertainty of 20\%.   The values of \citet{huc13} and this distance give ages of the nebular shells of 11,200 yrs
and 21,000 yrs.

To determine a reddening value we started with a weighted mean of the nebular Balmer decrement estimates from \citet{ack92}, \citet{kal83}, \citet{per94}, \citet{wal96}, and \citet{pol97} to determine $E(B-V)$ = 0.10. An independent value can be calculated from a comparison of the integrated 1.4\,GHz radio continuum and H$\alpha$ fluxes from \citet{con98} and \citet{fre13}, respectively; we do so to determine $E(B-V)$ = $0.13\pm0.06$.  Also, the asymptotic reddening on this sight-line from the revised reddening maps of \citet{sch11} is $E(B-V)$ = $0.09\pm0.03$.  However, a strong IR signature in the PN from dust found in WISE and IRAS data suggests a potentially strong nebular contribution to reddening.  This is confirmed by IUE LWP and SWP spectra.  The two spectra,
taken at different times show clearly different fluxes, but neither match the slope of a hot CS unless a reddening value $E(B-V)\gtrsim0.18$ is adopted.  Using the reddening model of \citet{val04a,val14b} we find the best match for $E(B-V)$ = 0.20 and adopt this as the reddening value for our modeling.

\begin{figure}[p]
\begin{center}
\includegraphics[width=6in]{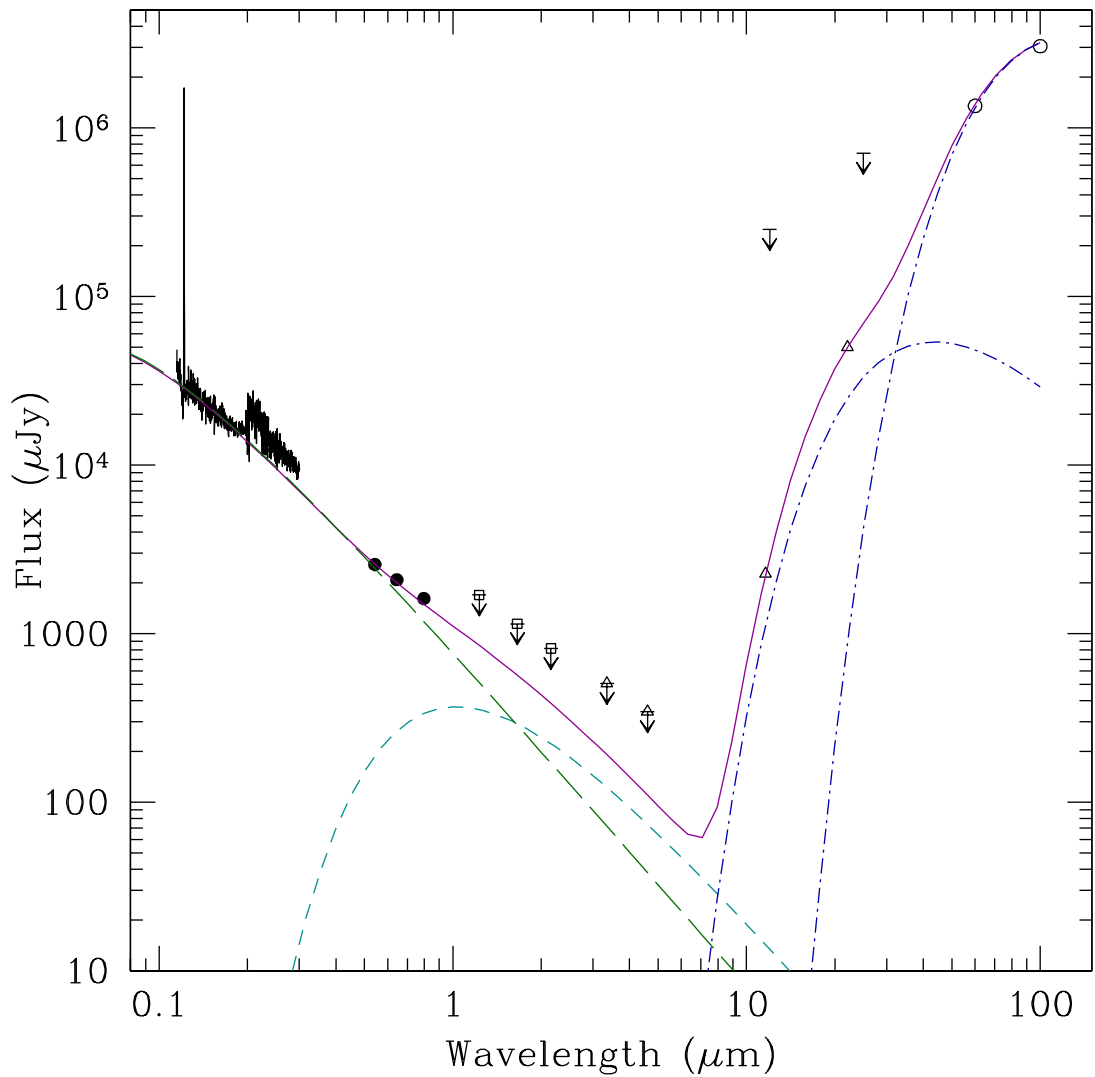}
\end{center}
\caption[Abell 65 SED] {The spectral energy distribution plot for the CS of Abell~65.  The IUE spectra, $V$, $R$,
and $I$ magnitudes ({\it solid circles}) at minimum light, 2MASS $J$, $H$, and $K$ magnitudes ({\it open squares}),
WISE fluxes ({\it open triangles}), and IRAS fluxes ({\it open circles}).  All fluxes have been
de-redenned assuming $E(B-V)=0.20$.  Also shown are model blackbody curves for our resulting CS, companion, and dust component, as well as the combined curve.  These are shown {\it without} any irradiation contribution, thus should approximately represent the system at minimum light.
\label{a65sed}}
\end{figure}

In Figure \ref{a65sed} we show the reddening corrected {\it IUE} spectra, $V$, $R$, and $I$ fluxes from this work, and {\it 2MASS} 
\citep{skr06}, {\it WISE}, and {\it IRAS} \citep{taj98} fluxes.  \citet{taj98} use corrected IRAS fluxes to determine a dust temperature
of 65 K.  Their correction
assumes extended dust emission over the entire optical nebula, so they scale the emission up from the size of the aperture to
the full size of the nebula.  Here we simply use the reported values, two of which are upper limits, as shown in the figure.

Using the ephemeris in \S4.1 we find that the SWP spectrum was obtained at phase 0.90, or close to
minimum light (which occurs at orbital phase $\phi=0$).  The LWP spectrum, obtained at $\phi=0.78$, has greater flux than the SWP spectrum, suggesting that part of the heated hemisphere of the companion is hot enough to affect the near-UV spectrum.  We find that this requires a temperature of at least 20,000 K on some portion of the companion.  Our final reddening value quoted above was found using the SWP spectrum along with our minimum light $V$ and $I$ values.  The remaining data from 2MASS, WISE, and IRAS were not obtained at minimum light so we do not expect them to match the included model, but they are shown to demonstrate the IR-excesses due to the companion and dust in the nebula.  In particular, the WISE and IRAS bands use large apertures and so
include both thermal dust emission and fine-structure line emission, as well as light from the central binary.  Indeed, in the WISE3 and WISE4 band images, the nebula overwhelms the star in brightness.  In the three IRAS bands, the magnitude difference between the observations and our stellar/dust model are 0.82, 0.99, and 1.09 magnitudes in the $J$, $H$, and $K$ bands respectively.  The
IRAS data were obtained at orbital phase 0.77 based on our ephemeris given below.  For that orbital phase, these values are
slightly higher than we would expect based on the irradiation effect alone.  Likewise, the WISE1 and WISE2 band values are 1.37 and 1.55 magnitudes above our minimum-light model, higher than expected from irradiation. However, these values are within a reasonable range when nebular contamination is also considered.  
We label these data points in Figure \ref{a65sed} as upper limits to remind the reader that due to nebular contamination and not being obtained at minimum light, they are upper limits for the model shown in the plot.
The WISE3 and WISE4 bands are used, along with the IRAS fluxes, to estimate the dust component of the system.

Without higher resolution in the infrared it is difficult to determine the source of the dust emission.  However, given that our best fitting reddening value is significantly higher than values from a number of other methods as described above, it seems likely that there may be an additional
compact dust source around the CS that contributes to the reddening, such as the dust disks reported for a number of PNe \citep[e.g.]{bil12}.  In Figure \ref{a65sed} we show a two cool dust components: a 115 K blackbody to fit the WISE3 and WISE4 bands, and 40 K blackbody fit to the IRAS 60 $\mu$m and 100 $\mu$m bands.
The cooler component roughly matches with the \citet{taj98} dust emission from the nebula, while the warmer component is
similar to dust disks seen in these systems, which average around 150 K.  In both cases, it would appear that there is an additional reddening contribution by dust in the nebula.

In \S 4 we further discuss several of the literature values for the Abell~65 nebula in the context of our binary star modeling.

\section{OBSERVATIONS AND REDUCTIONS}

The photometric data consist of images obtained over twenty years at four different telescopes in the $V$, $R$, and $I$ filters.  The Cerro Tololo Interamerican Observatory (CTIO) 0.9 meter and 1.3 meter telescopes, the SARA 0.9 meter telescope at Kitt Peak
National Observatory, and the Perth Observatory 24-inch telescope were all used to collect data.  The data include the
original discovery observations described in \citet{bon90} and follow-up observations by H. E. Bond on the CTIO 0.9 meter
telescope.  These data were collected in 1989 September, 1990 June, 1991 September, and 1992 May ($V$, $R$, and $I$ filters).  The SARA 0.9 meter telescope\footnote{Based on observations obtained with the SARA Observatory 0.9 m telescope at Kitt Peak, which is owned and operated by the Southeastern Association for Research in Astronomy (saraobservatory.org).}
was used to make observations from the northern hemisphere at multiple epochs in 2007 and 2008 ($R$ filter).  The CTIO 0.9 meter and Perth 24-inch telescopes were used to observe the CS of Abell 65 on three consecutive nights in 2008 August, providing coverage of  almost 70\% of the orbit each day ($V$, $R$, and $I$ filters).  The CTIO 1.3 meter provided observations in queue observing mode from August to October 2008 and June and July of 2009 ($R$ filter).  The Perth 24-inch telescope also provided observations in queue observing mode from 2008 June--August ($I$ filter).

All images were bias subtracted and flat fielded (and dark subtracted, when necessary).  The IRAF/DAOPHOT package was
used to perform aperture photometry on the images.  The 2007--2008 observations were then jointly analyzed using incomplete
inhomogeneous ensemble photometry \citep{hon92}.  The ensemble photometry used all of the stars in a template image to
perform differential photometry.  The large number of stars of various colors produces an effective color correction for slight
differences in system throughput between the four different telescope, filter, and camera combinations.  The success of this method
was confirmed in the final light curve by the seamless overlap of the 2007--2008 data from the four telescopes in the phase-folded light curves
(Figure \ref{a65phmod}).  The 1989--1992 data were analyzed by single star differential photometry with two check stars.  The
earlier data were then shifted to the same magnitude zero-point as the more recent data (see \S 4.1 below).  Some of these earlier
data do not align well with the 2007--2008 data, especially in the $V$ filter.  These differences could be a result of different
CCD and filter combinations, or could be intrinsic to the system (see \S 4.2 for more discussion).  The two data sets were combined to provide a more precise orbital ephemeris.  
  
The resulting light curves are in an instrumental magnitude system.
However, the photometric data were then shifted to apparent magnitudes using the $V = 15.80$ and $I = 15.39$ magnitudes from the HST observations of \citet{cia99}.  The final ephemeris for the binary system was used to determine the phase of the HST images (0.81 and 0.79 for $V$ and $I$, respectively) and our photometric data were shifted to those values at the determined phases based on our final binary model light curves.
Given the high precision of our ephemeris, we adopt a magnitude zero-point uncertainty of 0.06 mag based on the small uncertainty in orbital phase and the slope of the light curve at that phase.  To determine the shift necessary for our $R$ filter data, we
use a log fit between the $V$ and $I$ de-reddened fluxes to extrapolate an $R$ magnitude value for
Abell 65.  The resulting zero-point is thus slightly less well constrained, but should still be reliable to within 0.1 magnitudes.
The result gives apparent magnitudes at minimum light of $V = 16.04$, $R = 15.80$, and $I = 15.66$.  These minimum
brightness values provide the best observable measure of the combined brightness of the two stars in the system
with a minimum irradiation contribution.  They are used with the IUE SWP spectrum to arrive at a value for interstellar reddening, as
shown in Figure \ref{a65sed}.

The spectrum of the CS of Abell 65 was obtained on 2008 July 11 with the UVES high resolution spectrograph on the VLT \citep{dek00}.  The exposure time was 1200 s with a spectral resolution of $R\sim$60,000 and covering the range of 4800--6800 \AA .  
The spectrum was obtained at an airmass of 2.285 with the slit at the parallactic angle of 107$^\circ$.  The seeing was
1.38 arc seconds. The spectrum was reduced with the standard ESO UVES pipeline.  Details of the spectrum are discussed in \S 4.3.  

\section{ANALYSIS}

\subsection{Orbital Ephemeris}

Because we were unable to analyze the early epoch data using the same method as the more recent data (see \S 3 above) the two data sets had magnitude scales with different zero points.  To correct this a preliminary model of the binary system was produced using only the recent data with the Wilson-Devinney code \citep{wil90,wil71} in the $V$, $R$, and $I$ filters.  The resulting model was then fit to the early epoch data, allowing only the magnitude zero-point to change.
The derived zero-point offset was applied to the early epoch data to shift them to the same zero point as the more recent data.  Having all of the data on the same zero point scale allowed us to calculate the orbital ephemeris to our highest possible precision.  To do so we used the Wilson-Devinney code to fit the photometric data with only the time at phase $\phi = 0$, $T_0$, and the orbital period, $P_{orb}$, as variables.
The resulting binary system ephemeris is then $$T = 2454301.3244(38) + 1.0037577(10)\times\phi \label{eq1}$$
where $\phi = 0$ is taken to be at inferior conjunction of the cool secondary star (what would amount to primary eclipse if this were an eclipsing system).  This phase also corresponds to minimum light.  The values in parentheses reflect the uncertainties in the last two digits of each number.  The complete phase-folded $V$-, $R$-, and $I$-band light curves  are shown in Figure \ref{a65phmod}, where each has also been shifted to the apparent magnitude scale as described above.

\subsection{Binary System Parameters}

\citet{shi09} use their spectra and $V$-band light curve to calculate a model for the binary system confirming that it is indeed an irradiated binary with a hot central star and cool companion.  However, we believe that with our light curves in the $V$, $R$, and $I$ bands along with some additional information described below, we are able to produce a more accurate model of the system.  Because of the large number of parameters, binary system modeling often requires that a number of those parameters be estimated independently and used as fixed values in the modeling in order to reduce the degeneracy of the solution.  \citet{shi09} use the size and expansion rate of the PN to arrive at a nebular age and then use the evolutionary tracks of \citet{blo95} to derive a radius of the central star.  
The primary difference in our approach to modeling the system occurs at this point.  With our determined apparent magnitudes at minimum light, the published distances to Abell~65, and interstellar absorption values we determine the absolute magnitudes for the combined light from the binary CS of Abell~65.  We then constrain our models to match these values.

Using minimum light in the $V$ band ($V = 16.04$), the smallest distance value in the literature \citep[$d=1.17$ kpc,][]{fre08} and our adopted reddening (giving $A_V = 0.62$), we find M$_V = 5.09$ for the CS in Abell 65.
Using the largest value of distance, $d = 1.67$ kpc \citep*{sta08} we find an alternative minimum absolute magnitude for the CS of M$_V = 4.31$.  And with our adopted distance from \citet{fre14} of $d=1.4$ kpc, M$_V = 4.69$.

The binary parameters of \citet{shi09} result in an absolute magnitude, M$_V = 3.2$, well outside our range of $4.55\leq$ M$_V\leq5.33$.  It appears that the major cause of the discrepancy comes from their calculation of the age of the nebula (and thus the radius of the CS).  They quote a radius for the extended halo of Abell 65 from \citet{wal96} of 2 pc.  However, \citet{wal96} use a distance of 1.5 kpc to arrive at a {\it diameter} for the nebula of 2 pc rather than a radius.  The \citet{shi09} resulting age then is a factor of two too large.  Using an assumed expansion velocity and a larger distance than our adopted value of 1.4 kpc from \citet{fre08} leads to additional discrepancies in age.  The result is that their value for the age of the nebula of 130,000 years is considerably different than the result from \citet{huc13} of 21,000 years and 11,200 years for the outer and inner nebular shells, respectively (for our adopted distance).  Even with the difference in age it is not clear how \citet{shi09} arrive at their primary radius of 0.12 $R_\odot$.  Extrapolating between masses of 0.546 M$_\odot$ and 0.565 M$_\odot$ from the models of \citet{sch83} used by \citet{blo95} one
could find a value of the mass for which the CS radius would be approximately 0.12 $R_\odot$, but the basis for their claim that it is the most likely radius is not made clear.  However, their final model mass is 0.56 $M_\odot$ for which an age of 130,000 years places the CS much closer to the WD cooling track with a radius approximately a factor of five smaller.  It is largely due to the sensitive relationship between CS mass, radius, and age that we have added the absolute magnitude as a constraint on our models.  However, as discussed below we do use theoretical evolutionary tracks to help guide our modeling.

\begin{figure}[p]
\begin{center}
\includegraphics[width=6in]{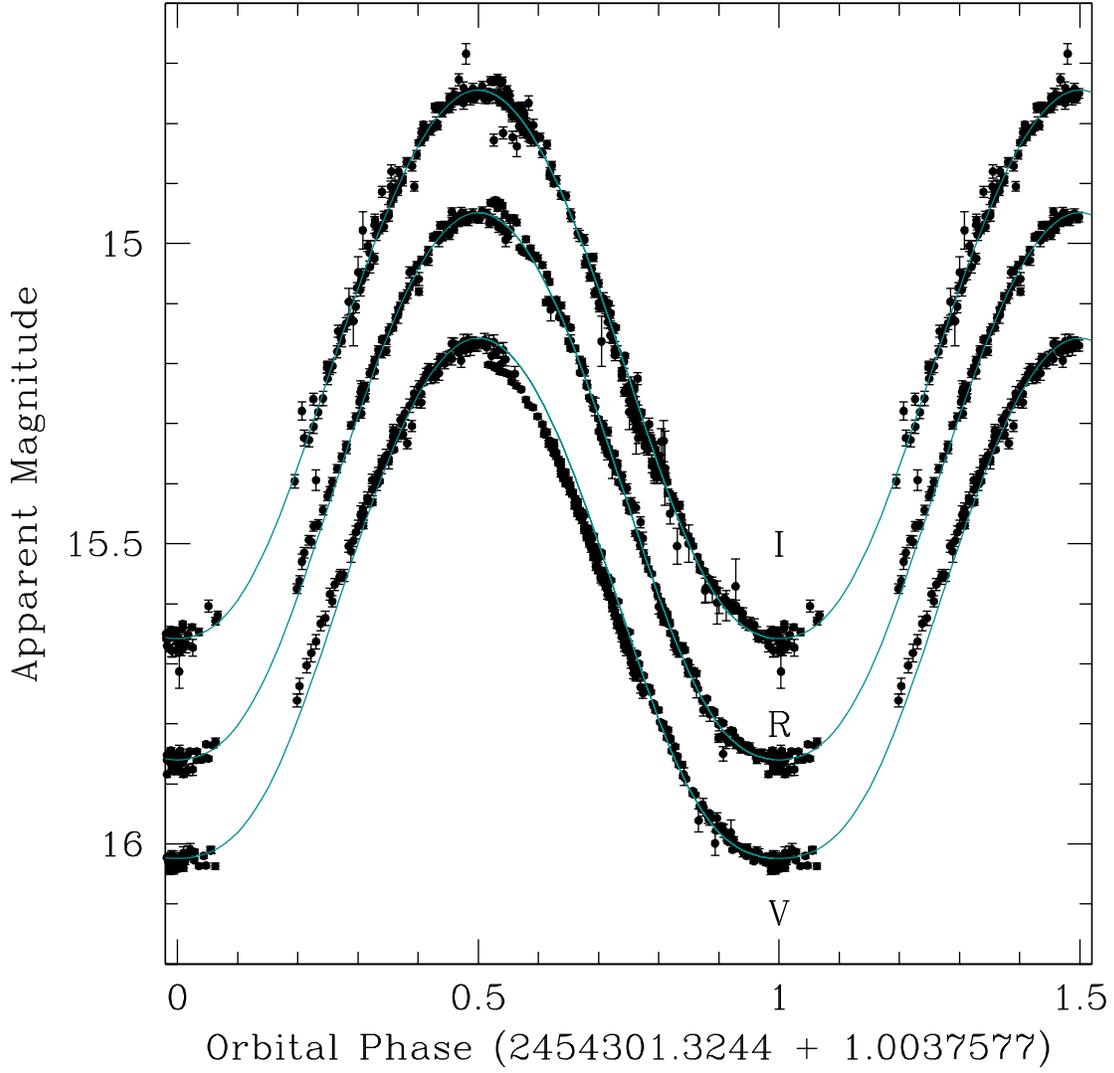}
\end{center}
\caption[Abell 65 Phase-Folded Light Curve] {The $V$, $R$, and $I$ phase-folded light curves
of Abell 65 for the period given in the ephemeris (Equation \ref{eq1}).  
\label{a65phmod}}
\end{figure}

Using the Wilson-Devinney code with our photometric data and the radial velocities of \citet{shi09} we find a best-fit solution for the system parameters.  In order to limit the possible parameter space we use the absolute magnitude constraint described above along with the temperature of the cool companion star found from \citet{shi09} from their spectra, $T_2 = 4950$ K.  The best-fit values from our modeling are given in Table \ref{a65model} and over-plotted with the data in Figure \ref{a65phmod}.  Because the uncertainties in fitting the data in binaries such as this are dominated by systematic effects, the uncertainties given for each parameter are not statistical errors for that individual parameter, but represent the range over which each parameter can vary in correlation with the other parameters.  The resulting uncertainties are larger than those found by holding all other parameters constant while varying only one, but provide a more physically significant range for each value.

In order to properly fit the light curves we departed from using, for the secondary star, the internally calculated limb-darkening
values in the Wilson-Devinney code.  We found that
lower values were necessary to reproduce the light curve shapes and amplitudes.  The model shown in Figure \ref{a65phmod} uses the internally calculated limb-darkening values for both
the $R$ and $I$ bands, but the $V$ band value has been reduced slightly.  The change in limb darkening required to fit the
observed light curves for any combination of parameter values in the ranges given in Table \ref{a65model} is within
10\% for all three bands.

We note several discrepancies in the photometry visible in Figure \ref{a65phmod}.  Both the $V$ and $R$ data on the increasing brightness side fall slightly above the model, though the effect is most evident in $V$.  The asymmetry is real and while most of the
data in that portion of the light curve are from a single night, there are a few data points from three different nights, suggesting
that the discrepancy is not instrumental or due to atmospheric conditions.  We find that we cannot match the discrepancy with our modeling.  The source of the asymmetry could potentially be due to non-synchronous rotation of the secondary and some of the
irradiated hemisphere rotating around toward the back hemisphere.  The greater effect with decreasing wavelength, the effect is
not noticeable in the $I$ data, would suggest an increase due to higher temperatures, as would be expected from asynchronous
rotation.  The orientation is also in the correct sense for prograde rotation
of the secondary at a rate greater than the orbital period.  There is no corresponding dip in the light curve below our model
on the decreasing brightness side of the light curve, as might be expected if synchronous rotation is moving the irradiated
portion of the atmosphere.  However, an asymmetry in that portion of the curve would only occur if the irradiation time scale
is on the order of the rotation time scale.  If the irradiation timescale is much shorter, then no such asymmetry would be expected.
We cannot at this point determine whether non-synchronous rotation
is the culprit, but it would seem to produce the observed features.  Another possible explanation would be a bright spot on the
cool companion, though a consistent bright spot at the same longitude on the star would seem unlikely.

We also find slight discrepancies in the peak of each light curve.  In $R$ and $I$ the data
were taken on different nights with the Perth telescope and the small brightness change appears to be inherent to the system.  Finally, the $V$ data on the
decreasing brightness branch from the early CTIO data (obtained in the 1990's) lie clearly below the model while the early $R$
data fall above the model.  The
$I$ data from the 1990's line up with the more recent data.  It is possible that
a difference in filter response is responsible.  Or there may be real changes in the amplitude over time.

The light curve amplitudes from our model are $\Delta V=0.882$ mag, $\Delta R=0.912$ mag, and $\Delta I=0.914$ mag.  The relative amplitude difference follows the relationship we presented in \citet{dem08}.

The resulting absolute magnitude of the system from our model values is M$_V=4.70$ which agrees well with the value found
above (M$_V=4.69$) using the distance of \citet{fre14}.  As discovered by \citet{shi09}, a hot and bright CS is required to fit the data, and we find that we are able to fit the data for relatively large ranges of temperature
and radius of both the CS and companion.  However, the degeneracy in the models is reduced dramatically by using an assumed system brightness.

In addition to our modeled absolute magnitude, several other system parameters differ from those of \citet{shi09}.  We find a companion with a lower mass, $M_2 = 0.22\pm0.04 M_{\odot}$ and radius, $R_2 = 0.41\pm0.05 R_\odot$ compared to their values
of $M_2 = 0.33\pm0.06 M_{\odot}$ and $R_2 = 0.59\pm0.08 R_\odot$, respectively.  
As with many other cool secondary companions in close binary CSPNe, we find that the secondary in Abell 65
has a higher temperature and radius than expected for a main sequence star of similar mass.  Thus the secondary is
{\it over-luminous} by about a factor of 20--30.  Our $M_2$ value suggests a companion spectral type M4V while the
temperature would be equivalent to a K2V star and the radius is appropriate for an M2-3V spectral type.  So both the
temperature and radius are large compared to those expected for a main sequence star based on our modeled mass.

The temperature of the CS of 110,000 K is larger than, but consistent with the Zanstra temperatures we calculate using the integrated H$\alpha$ flux from \citet{fre13} and the dereddened HeII/H$\alpha$ ratio averaged from the references given in Section 2.
The results are $T_{\rm z,H}$ = $49\pm7$ kK and $T_{\rm z,He}$ = $93\pm12$ kK (the large discrepancy between the hydrogen and ionized helium Zanstra temperatures indicates the nebula is optically thin).   It is also slightly higher than that of \citet{shi09}, but the two agree to within the uncertainties.  However, we find a CS radius more than a factor of 2 smaller than theirs.  As a consistency check we use our resulting $M_{CS}$, $R_{CS}$, log($T_{CS}$), and log($L_{CS}/L_\odot$) with the
evolutionary tracks of \citet{sch83} and \citet{blo95}.  The results are highly consistent with those tracks and lead to a CS age of about 22,000 years which, within the uncertainties in mass, radius, temperature, and distance, is in good agreement with the nebular age estimated by \citet{huc13} of $21,000\pm 7,000$ years (for the outer nebular shell, scaled to our distance).  The value of $11,200\pm 4,200$ years from \citet{huc13} for the inner shell is younger than the age we find from the evolutionary tracks.  This may be significant in that we would expect a CS that underwent CE evolution to evolve much faster than a non-CE star \citep[cf.][]{ric12,pas12}.  So if the CS in Abell~65 began its post-AGB evolution at a time consistent with ejection of the inner nebular shell, then we see a much shorter actual age than the predicted evolutionary age assuming a single star.  Because the models of \citet{huc13} are based on spatio-kinematic model we consider their ages to be the most reliable.  Even for the largest published distance for Abell~65 the age of the inner shell is still younger than the evolutionary age of the CS.  We do note however, that small deviations in the CS mass will
produce a large change in the evolutionary age, making this comparison of limited quantitative use given our uncertainty in $M_{CS}$.  In addition, the starting points for the evolutionary tracks are somewhat arbitrary and the spatio-kinematic models assume
no acceleration or deceleration of the expanding nebula.  So we again state that using this age comparison only serves as
a consistency check for our results, rather than an absolute comparison.

Our inclination, $i=61\pm 5\degr$, is lower than that of \citet{shi09} ($i=68.0\pm2.0\degr$), as a result of the different stellar parameters.  Our value though is in better agreement with the nebular inclination for the inner shell from \citet{huc13}, $i=55\pm 10\degr$, and also agrees with their value for the inclination of the outer shell, $i=68\pm 10\degr$.  As \citet{huc13} state, the connection between close binary CSPNe and their host nebula is becoming clearer.  For the handful of systems that have had both nebular and binary inclinations determined, the two values are within the uncertainties in every case: Abell~41 \citep{jon10}, Abell~63 \citep{mit07}, NGC~6337 
\citep{gar09,hil10}, HaTr 4 \citep[also Hillwig 2015a in prep]{tyn12}, NGC~6778 \citep{mis11,gue12}, Henize~2-428 \citep{san15}, and Sp~1 \citep[also Hillwig 2015b in prep]{jon12}.

\begin{center}
\begin{deluxetable}{lc}
\tablewidth{0pc}
\tablecolumns{3}
\tablecaption{Abell 65 Binary System Best-fit Model
\label{a65model}}
\tablehead{
\colhead {Parameter}        & \colhead{Value}}
\startdata
$q$						& $0.40\pm0.10$	\\
$T_{CS}$ ($\times10^3$ K) 	& $110 \pm10$ \\
$T_2$ ($\times10^3$ K) 		& $4.95\pm 1.0 $	\\
$M_{CS}$ ($M_\odot$) 		& $0.56 \pm0.04 $	\\
$M_2$ ($M_\odot$) 			& $0.22 \pm0.04 $	\\
$R_{CS}$ ($R_\odot$) 		& $ 0.056\pm0.008 $ \\
$R_2$ ($R_\odot$) 			& $ 0.41\pm0.05 $	\\
$\lg g_{CS}$ (cgs)			& $6.7\pm0.2 $		\\
$\lg g_2$ (cgs)				& $4.6\pm0.2  $ 	\\
$i$ ($\degr$) 				& $61 \pm 5$		\\
$a$ ($R_\odot$)  			& $3.90 \pm0.10 $	\\
log($L_{CS}/L_\odot$)		& $2.61$	\\
log($L_{2}/L_\odot$)			& $-1.05$	\\
\enddata
\end{deluxetable}
\end{center}

\subsection{The Spectrum of the Central Star of Abell 65}

\begin{figure}[p]
\begin{center}
\includegraphics[width=6in]{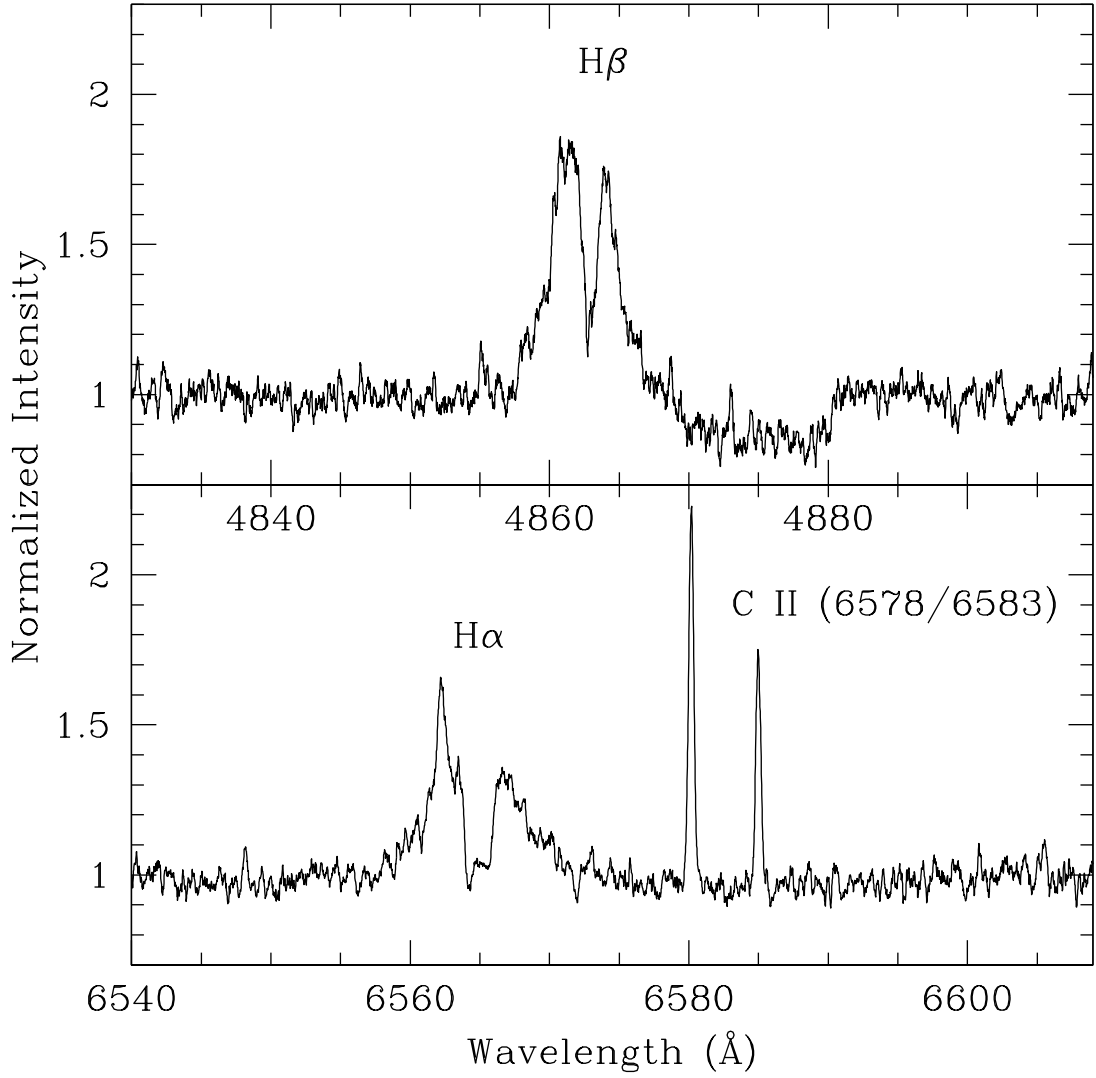}
\end{center}
\caption[Abell 65 Spectrum] {The normalized UVES/VLT spectrum of the CS of Abell 65 in the regions around H$\beta$ and H$\alpha$.  
\label{a65spectrum}}
\end{figure}

In Figure \ref{a65spectrum} we show a high resolution nebular subtracted spectrum of the CS of Abell~65 (see \S 3 for details) in the regions around H$\beta$ and H$\alpha$.  The spectrum was obtained at orbital phase $\phi=0.255$ based on our ephemeris.  What immediately stands out are the very broad \ion{H}{1} lines with central absorption.  These lines are reminiscent of those seen in the CS of HFG~1 \citep{ext05} with very broad double peaked \ion{H}{1} lines.  The line shape is reminiscent of the double-peaked \ion{H}{1} profiles of accretion disks in cataclysmic variables (CVs), thus the occasional classification of HFG~1 and Abell~65 as CVs.  However, this broad double-peaked line profile is well described for irradiated atmospheres by \citet{bar04} with the central absorption due to non-LTE effects in the upper atmosphere of the irradiated companion star (see their Figure 7).  The consistent change in equivalent width of these lines through the orbit \citep[see Figure 4 in][]{shi09} further confirms that the origin is not an accretion disk, for which the line strength would only change appreciably in the event of an eclipse.  Furthermore, the minimum light spectrum from \citet{shi09} shows the hot star to be
H-normal, giving an inferred spectral type of O(H).
The strong irradiation effect is also evidence of the non-CV nature of Abell~65 since for a CV the system brightness would be
dominated by the disk, thus swamping the irradiation effect and greatly reducing the observed variability amplitude.

\begin{figure}[p]
\begin{center}
\includegraphics[width=6in]{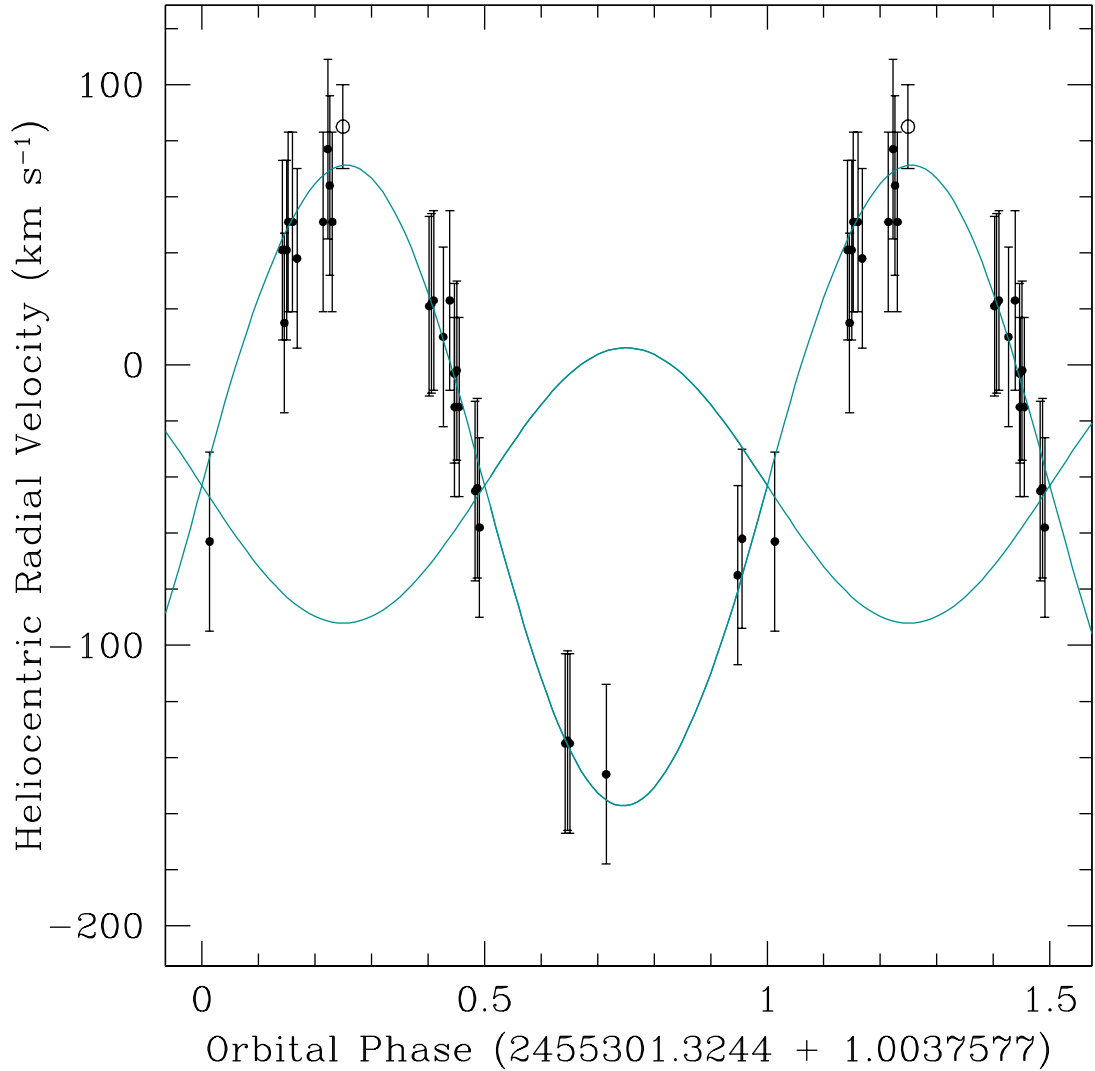}
\end{center}
\caption[Abell 65 Radial Velocity Plot] {The phase-folded radial velocity plot for the CS of Abell~65.  Shown are the
data from \citet{shi09} ({\it solid circles}), our single data point ({\it open circle}) from the VLT spectrum at phase 0.255, and our model fit ({\it solid lines}) which includes our predicted curve for the hot CS.
\label{a65rv}}
\end{figure}

We measured the radial velocity of the central absorption of the \ion{H}{1} lines in the VLT spectrum and the narrow \ion{C}{2} emission lines at 6578 and 6583 \AA, both of which originate from the secondary star.  The central absorption in \ion{H}{1} could be effected by incomplete or over subtraction of the nebulosity, but we find radial velocity values consistent with the narrow \ion{C}{2} lines, suggesting that any nebular subtraction effect is small.  The resulting velocity point is plotted in Figure \ref{a65rv} with the data of \citet{shi09} and our best-fit model parameters.  The VLT radial velocity point agrees well with both the previous data and our model.

\section{CONCLUSION}

\citet{shi09} have shown conclusively that the CS of Abell~65 is an irradiated close binary CSPN.  Due to the large parameter space available for binary system modeling, determining a final parameter set typically requires more information than radial
velocity and light curves.  With our three-color photometry we find that the parameters of \citet{shi09} do fit well.  However, we
believe that our use of luminosity as an initial fitting requirement results in a more accurate and more consistent parameter set
than their initial requirement of CS radius based on nebular age and CS mass.  Our results do not
change the overall identification of the system as a hot central star and cool, irradiated companion.  However, we believe that
our final set of physical parameters more accurately reflects the make-up of the binary system.  And we believe that what may seem relatively small differences in some of these parameters are important for future interpretations of the increasing body of close binary CSPNe.

Obtaining precise physical parameters for these systems can allow us to explore several interesting areas.  First, we find that in cases where both the nebular inclination and binary inclination are known, the values show a very strong correlation.  As the number of systems with known inclinations increases we will be able to say more about the likelihood that close binary interactions dominate the shaping mechanism in the resulting PN.  Second, as we increase the statistical significance of the close binary CSPN sample we can begin to explore the birthrates of cataclysmic variables (CVs) and type Ia supernova progenitors of various types.  Based on our results, the binary CS of Abell~65 will reach the semi-detached stage at a period of approximately 0.29 days and become a CV in approximately $9\times10^8$ years due to magnetic braking, based on calculations in \citet{war95}.  The time to a possible single or double degenerate Type Ia SN is longer than a Hubble time, as is expected for systems with periods of more than about half a day.  

As is the case for the large majority of post-CE CSs, this binary CS in Abell 65 has a period smaller than three days. Common envelope simulations \citep[e.g.,][]{ric12, pas12} predict a broader range of post-CE binary separations than is observed. It is this type of comparison that has alerted us to the fact that our model of the CE phase is far from complete. The irradiation properties of the companions in post-CE binaries are also going to serve as constraints of the CE interaction. Here too, CSs of PNe are unique, because the bright and hot primaries result in the brightest irradiation effects. One parameter that may be constrained is the amount of accreted angular momentum on the companion. Irradiation theory should be able to tell us whether the companion is synchronized or not, and hence give us information on its rotation properties. These properties connect to both its magnetic field, which is currently held responsible for bright and hard X-ray emission \citep[e.g.,][]{mon10, kas12} and the amount of accretion that may take place before or during the CE \citep*[e.g.,][]{mis13}.  Finally, as with most other secondary stars in close binary CSPNe, the secondary star here is both heated and expanded, thus over-luminous, compared to a main sequence star of similar mass.  This characteristic of close binary CSPNe is likely a result of the CE phase, though it is not clear why the stars have not returned to thermal equilibrium.  Increasing the statistical sample of over-luminous secondaries will hopefully lead to a better understanding of the CE and
post-CE phase effects on the companion star.

\acknowledgements

This material is based upon work supported by the National Science Foundation under Grant No. AST-1109683.  Any opinions, findings, and conclusions or recommendations expressed in this material are those of the author(s) and do not necessarily reflect the views of the National Science Foundation.  This publication makes use of data products from the Wide-field Infrared Survey Explorer, which is a joint project of the University of California, Los Angeles, and the Jet Propulsion Laboratory/California Institute of Technology, funded by the National Aeronautics and Space Administration.  This publication makes use of data products from the Two Micron All Sky Survey, which is a joint project of the University of Massachusetts and the Infrared Processing and Analysis Center/California Institute of Technology, funded by the National Aeronautics and Space Administration and the National Science Foundation.

Based on observations made with ESO Telescopes at the La Silla Paranal 
Observatory under programme ID 081.D-0857(A).

%\vfill\eject

%\clearpage

% Figure 1
%\figcaption {Light curve for NGC~6026 covering
%the CTIO 0.9m April/May 2002 data by night.\label{6026lc}}

% Figure 2
%\figcaption {Light curve for NGC~6337 covering the CTIO 0.9m April/May
%2002 data by night.\label{6337lc}}

% Figure 3
%\figcaption {Periodogram for the light curve of NGC~6026 in Figure
%\ref{6026lc}. The maximum power peak at 0.263 d is believed to
%represent the true orbital period.\label{6026pgram}}

% Figure 4
%\figcaption {Periodogram for the light curve of NGC~6337 in Figure
%\ref{6337lc}. The maximum power peak at 0.173 d is believed to
%represent the true orbital period.\label{6337pgram}}

% Figure 5
%\figcaption {The phase-folded light curve of NGC~6026 for the given
%ephemeris (Equation \ref{eq3}).\label{6026phase}}

% Figure 6
%\figcaption {The phase-folded light curve of NGC~6337 for the given
%ephemeris (Equation \ref{eq4}).\label{6337phase}}

%\clearpage
\pagestyle{empty}

%\clearpage
\pagestyle{empty}

\end{document}